# Magnetic geometry induced quantum geometry and nonlinear transports

Haiyuan Zhu[1, #], Jiayu Li[1,2, #], Xiaobing Chen[3,1], Yutong Yu[1], and Qihang Liu[1,3,4,*]

*[1]Department of Physics, State key laboratory of quantum functional materials, and Guangdong Basic Research Center of Excellence for Quantum Science, Southern University of Science and Technology, Shenzhen 518055, China*

*[2]New Cornerstone Science Laboratory, Department of Physics, The University of Hong Kong, Hong Kong 999077, China*

*[3]Quantum Science Center of Guangdong–Hong Kong–Macao Greater Bay Area (Guangdong), Shenzhen 518045, China*

*[4]Guangdong Provincial Key Laboratory for Computational Science and Material Design, Southern University of Science and Technology, Shenzhen 518055, China*

[#]These authors contributed equally: Haiyuan Zhu, Jiayu Li

[*]Email: liuqh@sustech.edu.cn

## Abstract

**The combination of quantum geometry and magnetic geometry in magnets excites diverse phenomena, some critical for antiferromagnetic spintronics. However, very few material platforms have been predicted and experimentally verified to date, with the material pool restricted by the assumed need for strong spin-orbit coupling (SOC). Here, we bypass the need for SOC by considering magnetic order induced quantum geometry and corresponding nonlinear transports (NLTs) in antiferromagnets (AFMs). By integrating spin space group theory into the symmetry analysis, we find that collinear and coplanar magnetic geometries can only induce NLT driven by Berry curvature dipole, and noncoplanar ones may trigger NLT driven by dipoles of Berry curvature, inverse mass, and quantum metric. Using this approach, we establish a materials database of 260 AFMs with SOC-free NLT effects, and complement this with first-principles calculations on several prototypical material candidates. Our work not only provides a universal theoretical framework for studying various magnetism-driven transport effects, but also predicts broad, experimentally accessible material platforms for antiferromagnetic spintronics.**

## Introduction

Nonlinear effects are widespread in various fields of modern physics, spanning from second harmonic generation in optics[1] to chaos in classical and quantum dynamics[2]. Spotlighting condensed matter systems, electrical nonlinear transport (NLT) is not only the foundation of next-generation devices such as full-wave rectification[3,4], but also a generic method to measure the distribution of the quantum geometry of states in momentum space[5-12]. In crystals, the quantum geometry, including Berry curvature and quantum metric, characterizes the curving and distances between neighboring Bloch states and is closely related to the topological properties of the system[13]. In magnetic crystals, the crossover between the quantum geometry in momentum space with the magnetic geometry in real space yields a diverse range of phenomena. Recently, it has been pointed out that second-order transports allow for the efficient detection of the Néel vector orientation making them ideal for antiferromagnetic spintronics[14-18].

Despite the promising applications, however, very few antiferromagnets (AFMs) have been theoretically predicted[9,10,17-21] and experimentally proven[11,12,22] to generate NLT so far. One reason is the long-known assumption that quantum geometry, accompanied by band anti-crossings, originates from spin-orbit coupling (SOC)[23-25], where the material pool was narrowed down to magnets composed of heavy elements. Another reason is that magnetic materials posess more complex and fragile magnetic structures, especially for AFMs. Experiments may observe results inconsistent with theoretical predictions due to the discrepancy of magnetic geometry[26-28].

To circumvent these issues we consider the NLTs[5,9,10,21,29-31] driven by magnetic geometry rather than SOC, termed geometric NLTs, in AFMs with experimentally verified magnetic configurations. It has been pointed out that complex magnetic geometry can inherently produce anomalous Hall effect[32,33], spin splitting[34-39], and spin-resolved transports[40-45]. Nevertheless, it is unclear for NLT whether magnetic geometry can trigger quantum geometry without the assistance of SOC. If so, it may generate more significant NLT due to the strong exchange interactions. Unfortunately, the conventional framework, where the NLT tensors are constrained by the magnetic space group of the magnetic geometry[10,46,47], provides no answer to this question. In magnetic space groups, rotational operations of spin and lattice are completely locked, thus entangling the magnetic geometry and SOC contributions to any effect. As such, predicting geometric NLTs relies on extensive computations *post factum*.

Here, we propose an efficient, symmetry-based framework to uncover the link connecting magnetic geometry and quantum geometry, and establish a database of inversion ($P$) broken AFMs with geometric second-order transports. We employ the spin space group (SSG) theory[43,48-55] to investigate the contributions of magnetic geometry toward quantum geometry and NLT. Our central result is that magnetic geometry generally triggers the quantum geometry and the second-order transports unless the effective symmetry suppresses all the components, as listed in Table I. By our framework, we deduce that collinear and coplanar magnetic geometry can only produce effects contributed by the Berry curvature dipole (BCD), where the combined symmetry of time-reversal ($T$) and spin rotation serves as the effective time-reversal $T_{\text{eff}}$ (see Fig. 1**a-b**) to eliminate quantum metric dipole (QMD) and inversed mass dipole (IMD). In contrast, noncoplanar magnetic geometry may, in general, produce all geometry quantities for both the longitudinal and transversal second-order transport once $T_{\text{eff}}$ and $PT_{\text{eff}}$ symmetries are absent (see Table I).

Within our SSG framework, we *a priori* single out 260 experimentally verified AFMs (120 collinear, 71 coplanar, and 69 noncoplanar magnetic configurations) from the MAGNDATA database[56,57] with geometric NLT. To demonstrate the accuracy of our framework, we present specific material candidates with density functional theory (DFT) calculations, including a collinear AFM $VNb_3S_6$ with $T_{\text{eff}}$ exhibiting NLT induced by BCD, a *room-temperature* noncoplanar AFM CrSe with $PT_{\text{eff}}$ exhibiting NLT induced by QMD, and other materials such as coplanar AFM $Ca_2Cr_2O_5$, noncoplanar AFMs, $CuB_2O_4$ and strain-engineered $Mn_3CoGe$, with NLTs driven by quantum geometric dipoles. Remarkably, we find that the magnitudes of geometric NLT can be comparable to or even larger than NLTs triggered by SOC, providing a deeper understanding of large NLTs in magnetic materials. Our work not only broadens the scope from magnetic geometry induced spin splitting to quantum geometry, but also paves a new avenue for the material discovery of nonlinear physics in unconventional magnets[58].

### Effective time-reversal symmetry in antiferromagnets

Let us first describe how the effective symmetries relevant to NLT emerged without SOC. Despite the breaking of $T$ in magnets, magnetic geometry may emerge effective time-reversal symmetry[59] $T_{\text{eff}}$ to constrain the NLT, where a well-known example is the combined symmetry of time-reversal and fractional lattice translation in the so-called $T\boldsymbol{\tau}$-AFMs, *e.g.* $MnBi_2Te_4$. More importantly, the magnetic geometry without SOC proceeds richer $T_{\mathrm{eff}}$ symmetry as the spin and lattice space are partially decoupled. Indeed, all the symmetries of the magnetic geometry form a SSG (Supplementary Section 1), where each symmetry operation takes the form $\{u||r|\boldsymbol{\tau}\}$ with $u$ and $r$ are the spin and lattice rotation, respectively, and $\boldsymbol{\tau}$ the lattice translation. Notice that the role of $T$ in spin space is analogous to that of inversion $P$ in lattice space. Then the key point is that without SOC, the charge transports is blind of proper spin rotation but affected by the *improper* one. Therefore, in any collinear AFM, the improper spin rotation $u = U_{\perp}(\pi)T$ maintains the magnetic geometry and serves as the $T_{\mathrm{eff}}$ symmetry (Fig. 1**a**), provided $U_{\perp}(\pi)$ a two-fold spinful rotation about an axis perpendicular to the Néel vector. Similarly, $T_{\mathrm{eff}}$ also emerges in any coplanar AFM since $u = U'_{\perp}(\pi)T$ always exists (Fig. 1**b**), with $U'_{\perp}(\pi)$ the two-fold spinful rotation along the axis normal to all magnetic moments. On the contrary, noncoplanar AFMs do not respect spin-only rotational symmetry, while some of them contain $T\boldsymbol{\tau}$ as $T_{\mathrm{eff}}$ (Fig. 1**c**). These effective time-reversal symmetries of the magnetic geometry constrain the $T$-odd charge transport tensors.

Besides $T$ symmetry, the combination of spatial inversion and time-reversal, $PT$, is also a crucial symmetry for NLT, as it suppresses the nonlinear Hall effect[5], where the collinear AFM CuMnAs is a famous instance[9,18] (Fig. 1**d**). For coplanar and noncoplanar AFMs, however, the exact $PT$ symmetry is generally missing owing to the complex magnetic geometry. Nevertheless, the absence of SOC allows the magnetic geometry to carry out the combined symmetry of improper spin rotation and spatial inversion as the effective $PT_{\mathrm{eff}}$. For example, a coplanar AFM shown in Fig. 1**e** is invariant under spatial inversion $P$ followed by spin rotation $U_y(\pi)T$ with $U_y(\pi)$ the two-fold spin rotation along the $y$ axis, and so does the noncoplanar AFM with $U_z(\pi)TP$, as presented in Fig. 1**f**. These $PT_{\mathrm{eff}}$ symmetries emerged from magnetic geometry constrain the $PT$-odd charge transport tensors.

**Second-order transport tensor and their symmetry constraint**

With the crucial effective symmetries of magnetic geometry, we next consider the NLTs, especially the second-order transports, originated from distinct geometric quantities. In general, the current density $\mathbf{J}$ driven quadratically by electric field $\mathbf{E}$ is given by $J^{\alpha} = \sigma^{\alpha\beta\gamma}E^{\beta}E^{\gamma}$ (Fig. 2**a**), where $\sigma^{\alpha\beta\gamma}$ is the second-order conductivity tensor of rank-three with spatial indices $\alpha, \beta, \gamma = x, y, z$, and the summation over repeated indices is implied. Finite second-order conductivity demands necessarily the breaking of $P$, resulting in dipole terms to generate NLT. Using the quantum kinetic theory[30,60] in weak scattering regime (ignoring disorder effects[61,62]), we deduce that three dipole terms contribute to the conductivity tensor with different polynomial dependences on relaxation time[20] ($\tau_r$): IMD contribution $\sigma_{\text{IMD}} \propto \tau_r^2$; BCD contribution $\sigma_{\text{BCD}} \propto \tau_r^1$ ; QMD contribution $\sigma_{\text{QMD}} \propto \tau_r^0$ (see Methods and Supplementary Section 2). The dependence on relaxation time encodes the symmetry transformation from distinct contributions under $T$ and $PT$: IMD and QMD contributions are $T$-odd, while BCD contribution is $PT$-odd (Methods). All these dipole terms have geometric significances. Specifically, the inverse mass tensor[18] $w_l^{\alpha\beta} = \hbar^{-2}\partial_\alpha\partial_\beta\varepsilon_l$, known as a Hessian tensor with $\partial_\alpha \equiv \partial/\partial_{k_\alpha}$, describes the local curvature of the $l$-th energy band manifold $\varepsilon_l$. In the meanwhile, quantum metric $G_l^{\alpha\beta}$ and Berry curvature $\Omega_l^{\alpha\beta}$, forming the quantum geometry tensor[13] $Q_l^{\alpha\beta} = G_l^{\alpha\beta} - i\Omega_l^{\alpha\beta}/2 = \sum_{n(\neq l)} A_{ln}^{\alpha} A_{nl}^{\beta}$, depict the geometric properties of the $l$-th Bloch state $|u_l\rangle$, where $A_{ln}^{\alpha} = i\langle u_l|\partial_\alpha u_n\rangle$ is the interband Berry connection. In second-order conductivity, the quantum metric is normalized to $\mathcal{G}_l^{\alpha\beta} = \text{Re}\left[\sum_{n(\neq l)} A_{ln}^{\alpha} A_{nl}^{\beta}/(\varepsilon_l - \varepsilon_n)\right]$, also known as the Berry connection polarizability[10,29]. All three dipole terms can be considered as the electrons on the Fermi surface carrying special charges of $w_l^{\alpha\beta}$, $\Omega_l^{\alpha\beta}$, and $\mathcal{G}_l^{\alpha\beta}$, and transporting with group velocity $v_l^{\gamma}$, as shown in Fig. 2**b**-**d**.

We now employ symmetry analysis to restrict NLT tensors. First for general symmetries $P$ and $T$, all the dipole terms are $P$-odd and so does the second-order conductivity, while the conductivity tensors contributed by IMD and QMD are $T$-odd and that by BCD is $T$-even. Combining them yields that IMD and QMD contributions are $PT$-even and BCD contributions is $PT$-odd. These relations are also valid for effective symmetries of $T_{\text{eff}}$ and $PT_{\text{eff}}$ emerged in AFM, and the general symmetry constraint on NLT conductivities are collected in Table I. As one can see, IMD and QMD share the same symmetry constraint. Two consequences are concluded: (i) in noncentrosymmetric AFM with collinear and coplanar magnetic geometry, only the BCD-contributed non-longitudinal current is allowed if no $PT_{\text{eff}}$ emerges. (ii) only noncentrosymmetric AFM of noncoplanar magnetic geometry allow longitudinal and transversal current contributed by IMD and QMD if no $T_{\text{eff}}$ emerges, and the BCD contribution is additionally allowed if no $PT_{\text{eff}}$ exists. Note here that the allowance by $T_{\text{eff}}$ or $PT_{\text{eff}}$ does not necessarily imply the existence of NLT, as other spin group symmetry constraints still need to be considered. Given any spin group symmetry $\{u||r|\boldsymbol{\tau}\}$, the BCD and IMD/QMD tensors are transformed by

$$\sigma_{\mathrm{BCD}}^{\alpha\beta\gamma} = \mathcal{R}^{\alpha\mu}\mathcal{R}^{\beta\nu}\mathcal{R}^{\gamma\eta}\sigma_{\mathrm{BCD}}^{\mu\nu\eta}, \tag{1}$$

$$\sigma_{\mathrm{IMD/QMD}}^{\alpha\beta\gamma} = \det(\mathcal{U})\,\mathcal{R}^{\alpha\mu}\mathcal{R}^{\beta\nu}\mathcal{R}^{\gamma\eta}\sigma_{\mathrm{IMD/QMD}}^{\mu\nu\eta}, \tag{2}$$

where $\mathcal{R}$ and $\mathcal{U}$ are the representation matrices of $r$ and $u$ under Cartesian coordinate. Considering the BCD contributed tensors as vector $\{\sigma_{\mathrm{BCD}}^{xxx}, \cdots, \sigma_{\mathrm{BCD}}^{zzz}\}$, Eq. (1) provides the linear transformation of it and the eigenvectors of the transformation solves the $\{u||r|\boldsymbol{\tau}\}$-allowed BCD contributed tensors (see Supplementary Section 3), where the procedure for allowed tensors contributed by IMD/QMD is same. In nonmagnetic systems, the procedure for BCD leads us only gyrotropic point groups[63] except for the tetrahedral and octahedral groups are sufficient for nonzero BCD contributions (see Supplementary Section 4). Here for AFM, provided its SSG, the symmetry-allowed conductivity tensors of any magnetic geometry can be predicted by Eqs. (1) and (2). Notice that BCD contribution bears extra constraint of $\sigma_{\mathrm{BCD}}^{\alpha\beta\gamma} + \sigma_{\mathrm{BCD}}^{\beta\gamma\alpha} + \sigma_{\mathrm{BCD}}^{\gamma\alpha\beta} = 0$ (see Methods and Supplementary Section 3).

**Diagnosis of realistic materials**

To materialize the magnetic-geometry-induced NLT, we construct a complete database of validated AFMs with second-order conductivity tensors allowed by SSG symmetry. Starting from ~1700 experimentally validated AFMs in MAGNDATA database[56,57] on the Bilbao Crystallographic Server (BCS; http://www.cryst.ehu.es), we selected 803 noncentrosymmetric AFMs as material pool. Subsequently, the SSG of each AFM was recognized by our online program FINDSPINGROUP (https://findspingroup.com)[51]. With the SSG for any AFM at hand, we predicted which NLT contributions are possible by checking $T_{\mathrm{eff}}$ and $PT_{\mathrm{eff}}$ according to Table I, and further solved which tensor components are allowed under the constraints by SSG using Eqs. (1) and (2) implemented in FINDSPINGROUP. Finally, a database of 260 AFMs with geometric NLT tensors induced by magnetic geometry is established, where 120 collinear and 71 coplanar AFMs allow BCD-contributed NLT. We also found 69 noncoplanar AFMs with SSG-allowed NLT, 21 among which feature all NLT contributed by both IMD, BCD, and QMD. Our comprehensive database includes large fraction about 32.4% of the 803 noncentrosymmetric AFMs, revising the previous consensus that nontrivial transports are generally triggered by SOC.

A snapshot of the AFM database is presented in Fig. 2**e**, and the full list is provided in Supplementary Section 4, Table S1-S3 as well as the online database FINDSPINGROUP.

**Material examples**

Below we choose two candidates from our database and performed DFT level calculations on the second-order transport tensors (Methods). The first example is the transition metal $VNb_3S_6$ crystallized by 1H-$NbS_2$ layers with V inserted at interlayer positions as shown in Fig. 3**a**, where the magnetic moments in the same (adjacent) V layer are parallel (anti-parallel) with the Néel vector orientated along $a$ axis[64]. The SSG is recognized as $P^{-1}6_3{}^{-1}2^{1}2^{\infty m}1$, which is generated by spatial rotations $\{T||R_{[100]}(\pi)|(1/2)\boldsymbol{\tau}_c\}$, $\{E||R_{[120]}(\pi)|\mathbf{0}\}$, skew rotations $\{T||R_z(\pi/3)|(1/2)\boldsymbol{\tau}_c\}$ with $\boldsymbol{\tau}_c = (0,0,1)$ the lattice translation, and the spin-only subgroup ${}^{\infty m}1$ of infinite spin rotation along the Néel vector. The band structure without SOC, as shown in Fig. 3**b**, exhibits the SOC-free spin splitting due to the breaking of $T$. By the newly definition of altermagnetism, collinear AFM $VNb_3S_6$ is a so-called $g$-wave altermagnet[65]. By Table I, we predict that the $T_{\text{eff}}$ symmetry of $U_z(\pi)T$ naturally forbid the IMD/QMD contributions, while the absence of $PT_{\text{eff}}$ symmetry implies the BCD-contributed conductivity tensor to be allowed. From our database (Supplementary Section 4, Table S1), the allowed BCD-contributed tensor components of $VNb_3S_6$, constrained by SSG with Eq. (1), are $\sigma_{\text{BCD}}^{xyz} = -\sigma_{\text{BCD}}^{yxz}$. To verify this, the BCD-contributed tensor components are computed without SOC in Fig. 3**c**, showcasing that the quantum-geometry-driven NLT effect can be inherently induced by magnetic geometry. Moreover, the maximum value of $\sigma_{\text{BCD}}^{xyz}$ approaching $75\ \text{S}^2/\text{A}$ (with relaxation time set $\tau_r = 1$ ns) at $\sim 0.25$ eV above the Fermi energy, which is comparable to the nonlinear Hall conductivity of CuMnSb[17]. Such a large conductivity originates from the BCD hot spot (Fig. 3**d**) at the corresponding energy. Note that considering SOC barely changes the BCD contribution around the Fermi energy and only increases the peak by $\sim 6\%$ (Supplementary Section 5, Fig. S6), indicating that magnetic geometry is the dominate driving force of NLT in $VNb_3S_6$ even with SOC counted.

Our next example is CrSe of room temperature (Néel temperature $290\ \text{K}$) noncoplanar antiferromagnetic geometry[66,67], as shown in Fig. 4**a**. Separated by Se layers, Cr atoms form layers of trigonal sublattice in $ab$ plane, where the in-plane magnetic components inside each Cr layer are related by spin rotation $U_z(2\pi/3)$ with alternating out-of-plane (along $c$) magnetic components between neighboring layers. The SSG of CrSe is $P^{2_{010}}6_3/{}^{-1}m^{m_{010}}m^{-1}c|(3^2_{001}, 3^2_{001}, 1)$, which is generated by $\{U_{[010]}(\pi)||R_z(\pi/3)|(1/2)\boldsymbol{\tau}_c\}$, $\{U_{[010]}(\pi)T||P|\mathbf{0}\}$, $\{E||R_{[210]}(\pi)|\mathbf{0}\}$, and spin screw rotation $\{U_z(2\pi/3)||E|(1/3)\boldsymbol{\tau}_a + (2/3)\boldsymbol{\tau}_b\}$ and $\{U_z(-2\pi/3)||E|(2/3)\boldsymbol{\tau}_a + (1/3)\boldsymbol{\tau}_b\}$ with $\boldsymbol{\tau}_a = (1,0,0)$ and $\boldsymbol{\tau}_b = (0,1,0)$. We find that CrSe contains $\{U_{[010]}(\pi)T||P|\mathbf{0}\}$ as $PT_{\text{eff}}$ protecting the four-fold and six-fold band degeneracy at Γ and $K$, respectively[51] (Fig. 4**b**). Besides band degeneracy, $PT_{\text{eff}}$ further eliminates the BCD contribution of NLT as seen from Table I. By referring to the database (Supplementary Section 4, Table S3), we predict the SSG-allowed conductivity tensor components, contributed by QMD, are $\sigma^{xyz}_{\text{QMD}} = -\sigma^{yxz}_{\text{QMD}}$. Once again, our DFT calculations on QMD-contributed tensor components in Fig. 4**c** are consistent with the spin group analysis, where the maximum is $23.4\ \text{S}^2/\text{A}$ at $\sim$0.19 eV above Fermi energy, corresponding to the significant QMD (Fig. 4**d**). Such QMD contribution triggered by magnetic geometry is much larger than that of $\sim 0.01\ \text{S}^2/\text{A}$ in $MnBi_2Te_4$ thin film[12,21], which is purely triggered by SOC. We note that SOC induces opposite contributions on $\sigma^{xyz}_{\text{IMD}}$, resulting in a net value of $\sim 0.2\ \text{S}^2/\text{A}$ (Supplementary Section 6, Fig. S10). Such a large net NLT originates from the uncompensated contributions from both magnetic geometry and SOC, implying the overlooked contribution from magnetic geometry is not negligible in CrSe.

Moreover, we note here that even though the magnetic geometry and SOC contributions are entangled in realistic materials, one can quantitatively estimate the geometric NLT by well-designed devices and setups. For instance, in $VNb_3S_6$, we find that with the pure SOC contributions $\sigma^{zxy}_{\text{BCD}}$ measured in experiment, the geometric NLT of $\sigma^{xyz}_{\text{BCD}}$ can be quantitatively estimated from experimental data through $\sigma^{xyz}_{\text{BCD}}(\text{w/o SOC}) \approx \sigma^{xyz}_{\text{BCD}} - \sigma^{zxy}_{\text{BCD}}$, with $\sigma^{xyz}_{\text{BCD}}$ measured in the $(0001)$ surface and $\sigma^{zxy}_{\text{BCD}}$ in the $(10\bar{1}0)$ surface (Supplementary Section 5, Fig. S6). On top of using other component ($\sigma^{zxy}_{\text{BCD}}$), one can also experimentally evaluate the geometric NLT $\sigma^{xyz}_{\text{BCD}}$ itself by angular-resolved signals in a multiterminal device via the sum frequency method (Supplementary Section 7, Fig. S15).

Besides $VNb_3S_6$ and CrSe, we also perform DFT calculations on $Ca_2Cr_2O_5$[68], $CuB_2O_4$[69], and $Mn_3CoGe$[70], and the results are summarized as follows: the coplanar magnetic geometry of $Ca_2Cr_2O_5$ forbids the $T$-odd effects, leaving two independent BCD components to be finite; $CuB_2O_4$ with noncoplanar magnetic geometry allows both $T$-odd and $T$-even conductivity components due to the absence of $T_{\mathrm{eff}}$ and $PT_{\mathrm{eff}}$; $Mn_3CoGe$ with noncoplanar magnetic geometry should allow two independent $T$-odd and $T$-even conductivity components, *i.e.*, $\sigma^{xyz} = \sigma^{yzx} = \sigma^{zxy}$ and $\sigma^{yxz} = \sigma^{zyx} = \sigma^{xzy}$. However, the extra constraint on BCD components $\sigma_{\mathrm{BCD}}^{xyz} + \sigma_{\mathrm{BCD}}^{yzx} + \sigma_{\mathrm{BCD}}^{zxy} = 0$ eliminates all the BCD contributions, where a uniaxial strain $\epsilon_z$ breaks the three-fold rotation along $[111]$ direction and also the identity $\sigma^{xyz} = \sigma^{yzx} = \sigma^{zxy}$. Hence, uniaxial strain can induce BCD-contributed NLT in $Mn_3CoGe$. All these results are consistent with our *a priori* predictions in Supplementary Table S1-S3 and the details are provided in Supplementary Sections 5 and 6.

**Discussion**

We propose an efficient framework to search for AFMs with magnetic geometry driven quantum geometry and second-order charge transport. Our diagnosis of magnetic geometry triggered second-order transport is based on the complete symmetry analysis of SSG rather than the conventional magnetic space group. Within magnetic space group, the symmetry-allowed NLT tensors could be triggered by both of magnetic geometry and relativistic SOC. One can only disentangle their contributions *post factum* by performing time-consuming first-principles calculations on NLT tensors with and without SOC. Nevertheless, with the help of SSG, we *a priori* predict which NLT tensors are triggered by magnetic geometry for any AFM. After that, the SOC contributions can be immediately extracted by comparing the allowed NLT tensors constrained by the SSG and magnetic space group. For instance, we can directly point out that the experimentally observed NLT of $MnBi_2Te_4$ induced by QMD is a pure SOC effect[12,21]. Furthermore, our framework is universal for other magnetic-geometry-induced nonlinear effects like photovoltaic effects[71] and current-induced spin polarization[72,73]. It can also be easily extended for the third order transport effects[74-78], which could be the leading order of electrical transport effects in certain centrosymmetric magnets.

## Methods

### *Nonlinear charge transport*

In general, the current density **J** driven quadratically by electric field **E** is given by $J^{\alpha} = \sigma^{\alpha\beta\gamma}E^{\beta}E^{\gamma}$. Here $\sigma^{\alpha\beta\gamma}$ is the second-order conductivity tensor. The conductivity tensor can be derived within the quantum kinetic theory[30,60]. Here we concentrate on the weak scattering limit, *i.e.*, ignoring disorder contributions[61,62]. Under the relaxation time approximation, the dynamic of the density matrix is encoded by quantum Liouville-von Neumann equation:

$$\frac{i}{\hbar}\left[H_0, \rho^{(N)}\right]_{ln} + \frac{\rho_{ln}^{(N)}}{\tau_r/N} = -i\frac{e\mathbf{E}}{\hbar}\cdot\left[\mathbf{r}, \rho^{(N-1)}\right]_{ln}, \tag{3}$$

where $H_0$ is the field-free Hamiltonian, $\rho^{(N)} \propto E^N$ is the field-perturbated density matrix, $\tau_r$ is the relaxation time, $\mathbf{r}$ is the position operator, $l, n$ are the band indices. After tedious derivation, we obtain three distinct conductivity tensors contributed by IMD, BCD, and QMD, respectively, read

$$\sigma_{\mathrm{IMD}}^{\alpha\beta\gamma}(\tau_r^2) = \tau_r^2\frac{e^3}{2}\sum_{\mathbf{k},l} v_l^{\alpha} w_l^{\beta\gamma}\frac{\partial f_l}{\partial \varepsilon_l}, \tag{4}$$

$$\sigma_{\mathrm{BCD}}^{\alpha\beta\gamma}(\tau_r^1) = -\tau_r\frac{e^3}{2\hbar}\sum_{\mathbf{k},l}\left(v_l^{\gamma}\Omega_l^{\alpha\beta} + v_l^{\beta}\Omega_l^{\alpha\gamma}\right)\frac{\partial f_l}{\partial \varepsilon_l}, \tag{5}$$

$$\sigma_{\mathrm{QMD}}^{\alpha\beta\gamma}(\tau_r^0) = e^3\sum_{\mathbf{k},l}\left[v_l^{\alpha}\mathcal{G}_l^{\beta\gamma} - 2\left(v_l^{\gamma}\mathcal{G}_l^{\alpha\beta} + v_l^{\beta}\mathcal{G}_l^{\alpha\gamma}\right)\right]\frac{\partial f_l}{\partial \varepsilon_l}, \tag{6}$$

Here $w_l^{\alpha\beta} = \hbar^{-2}\partial_{\alpha}\partial_{\beta}\varepsilon_l$ is the inverse mass tensor, $\Omega_l^{\alpha\beta} = -2\mathrm{Im}\left[\sum_{n(\neq l)} A_{ln}^{\alpha}A_{nl}^{\beta}\right]$ is the Berry curvature tensor, $\mathcal{G}_l^{\alpha\beta} = \mathrm{Re}\left[\sum_{n(\neq l)} A_{ln}^{\alpha}A_{nl}^{\beta}/(\varepsilon_l - \varepsilon_n)\right]$ is the band-normalized quantum metric tensor (also called Berry connection polarizability), $v_l^{\gamma}$ is the group velocity of $l$-th band $\varepsilon_l$ of $\gamma$ component, and $f_l = \{1 + \exp[(\varepsilon_l - \mu)/k_B T]\}^{-1}$ is the Fermi distribution function of band $\varepsilon_l$. For full derivation, please see Supplementary Section 2.

As IMD and QMD contributions are proportional to the even order of $\tau_r$, they are odd function under time-reversal $T$ but even function under the combination of spatial inversion and time-reversal $PT$. On the contrary, the BCD contribution proportional to the odd order of $\tau_r$ is $T$-even but $PT$-odd. This can be seen by expanding $\sigma^{\alpha\beta\gamma}$ with respect to $\tau_r$ as $\sigma^{\alpha\beta\gamma} = \sum_i \tau_r^i \chi_i^{\alpha\beta\gamma}$ . Since time-reversal operator maps $\sigma^{\alpha\beta\gamma} \to -\sigma^{\alpha\beta\gamma}$ and $\tau_r \to -\tau_r$, $\chi_{2n}^{\alpha\beta\gamma}$ is an odd function of $T$ and $\chi_{2n+1}^{\alpha\beta\gamma}$ is an even function. Thus $\sigma_{\mathrm{IMD}}^{\alpha\beta\gamma} \propto \chi_2^{\alpha\beta\gamma}$ and $\sigma_{\mathrm{QMD}}^{\alpha\beta\gamma} \propto \chi_0^{\alpha\beta\gamma}$ are $T$-odd, while $\sigma_{\mathrm{BCD}}^{\alpha\beta\gamma} \propto \chi_1^{\alpha\beta\gamma}$ is $T$-even. As spatial inversion $P$ flips $\sigma^{\alpha\beta\gamma} \to -\sigma^{\alpha\beta\gamma}$ but preserves $\tau_r$, all the three contributions are $P$-odd, implying that IMD and QMD contributions are $PT$-even but BCD contribution is $PT$-odd.

Moreover, we note that $\sigma_{\mathrm{IMD}}^{\alpha\beta\gamma}$ is symmetric under any permutation of all three indices $\alpha$, $\beta$, and $\gamma$, where $\sigma_{\mathrm{BCD}}^{\alpha\beta\gamma}$ and $\sigma_{\mathrm{QMD}}^{\alpha\beta\gamma}$ are symmetric under permutation of the last two indices $\beta \leftrightarrow \gamma$. Moreover, the BCD contribution bears one extra constraint[79]:

$$\sigma_{\mathrm{BCD}}^{\alpha\beta\gamma} + \sigma_{\mathrm{BCD}}^{\beta\gamma\alpha} + \sigma_{\mathrm{BCD}}^{\gamma\alpha\beta} = 0. \tag{7}$$

This simple relation implies three explicit constraints:

$$\begin{cases} \sigma_{\mathrm{BCD}}^{xyz} + \sigma_{\mathrm{BCD}}^{yzx} + \sigma_{\mathrm{BCD}}^{zxy} = 0, & \alpha \neq \beta \neq \gamma \neq \alpha, \\ \sigma_{\mathrm{BCD}}^{\alpha\alpha\gamma} = \sigma_{\mathrm{BCD}}^{\alpha\gamma\alpha} = -\dfrac{1}{2}\sigma_{\mathrm{BCD}}^{\gamma\alpha\alpha}, & \alpha = \beta \neq \gamma, \\ \sigma_{\mathrm{BCD}}^{\alpha\alpha\alpha} = 0, & \alpha = \beta = \gamma. \end{cases} \tag{8}$$

The last constraint implies the non-longitudinal nature of the BCD contribution, which is ensured by the anti-symmetry of Berry curvature $\Omega_l^{\alpha\beta}(\mathbf{k}) = -\Omega_l^{\beta\alpha}(\mathbf{k})$ and $\sigma_{\mathrm{BCD}}^{\alpha\alpha\alpha} \propto v_l^{\alpha}\Omega_l^{\alpha\alpha} = 0$.

***Density functional theory calculations.***

All DFT calculations herein are performed using projector augmented wave method[80], implemented in Vienna ab initio simulation package (VASP)[81]. The generalized gradient approximation of the Perdew-Burke-Ernzerhof-type exchange-correlation potential[82] is adopted. All the verified AFMs are collected in the MAGNDATA database (https://www.cryst.ehu.es/magndata/). For 20-atom collinear AFM $VNb_3S_6$, #0.712 in MAGNDATA with lattice parameter of the magnetic unit cell $a = b = 5.73$ Å, $c = 12.11$ Å, we solve (3p,4s,3d) electrons for V, (4p,5s,4d) electrons for Nb, (3s, 3p) electrons for S, with $E_{\mathrm{cut}} = 400$eV and a k-point mesh of $13 \times 13 \times 5$. For 12-atom non-coplanar AFM CrSe, #2.35 in MAGNDATA with lattice parameter of the magnetic unit cell $a = b = 6.37$ Å, $c = 6.02$ Å, we solve (3p,4s,3d) electrons for Cr, (4s,4p) electrons for Se, with $E_{\mathrm{cut}} = 500$ eV and a k-point mesh of $13 \times 13 \times 9$. Tight-binding models are constructed from DFT bands using the WANNIER90 package[83], and NLT tensors in gauge-covariant form[79] are calculated within WannierBerri code[84]. The numbers of sampled k-points in Fig. 3**c** and Fig. 4**c** are 1100*1100*1100 and 1700*1700*1700, respectively. Crystal structures are plotted by VESTA[85]. The SSGs of materials, labeled by the international notation[51], are diagnosed by the self-developed program FINDSPINGROUP at https://findspingroup.com.

**Data availability**

The data of the SSGs and tensors of AFMs are available with an interactive user interface at https://findspingroup.com. The supportive data for plots and other fundings in this study are available from the corresponding author upon reasonable request.

**Code availability**

The computation code for getting the theoretical prediction is available from the corresponding authors upon reasonable request.

**Acknowledgements**

We thank Weizhao Chen, Hai-Zhou Lu, and Xiaoxiong Liu for the helpful discussions and assistance in WannierBerri codes. This work was supported by National Key R&D Program of China under Grant No. 2020YFA0308900, National Natural Science Foundation of China under Grant No. 12274194, Guangdong Provincial Quantum Science Strategic Initiative (Grant No. GDZX2401002 and No. GDZX2201001), Guangdong Provincial Key Laboratory for Computational Science and Material Design under Grant No. 2019B030301001, Shenzhen Science and Technology Program (Grant No. RCJC20221008092722009 and No. 20231117091158001), the Innovative Team of General Higher Educational Institutes in Guangdong Province (Grant No. 2020KCXTD001) and Center for Computational Science and Engineering of Southern University of Science and Technology.

**Author contributions**

Q.L. conceived the project. H.Z. and J.L. performed symmetry analysis and DFT calculations on nonlinear transport tensors in antiferromagnets and established the database. J.L. developed the gauge-covariant form of QMD contribution based on WannierBerri package. X.C. and Y.Y. identified spin space groups of antiferromagnets in MAGNDATA. Y.Y. built the tensor part on the website. J.L., H.Z., and Q.L. prepared the manuscript with contributions from all authors. H.Z. and J.L. contributed equally to this work.

**Competing interests**

The authors declare no competing interests.

**Table I. | Magnetic geometry induced second-order conductivities by symmetry analysis**. All the AFMs are supposed to be noncentrosymmetric. In the column of $T_{\mathrm{eff}}$ and $PT_{\mathrm{eff}}$, ✔ and ✘ represent the presence and absence of the specific effective symmetry, respectively. In the column of "IMD & QMD" and "BCD", ✔ and ✘ denote the corresponding nonlinear transport to be allowed and disallowed by $T_{\mathrm{eff}}$ and/or $PT_{\mathrm{eff}}$, respectively.

| AFM | $T_{\mathrm{eff}}$ | $PT_{\mathrm{eff}}$ | IMD & QMD | BCD | Representative compounds |
|---|---|---|---|---|---|
| Coplanar (include collinear) | ✔ | ✔ | ✘ | ✘ | $MnBi_2Te_4$ |
| | ✔ | ✘ | ✘ | ✔ | $VNb_3S_6$ |
| Noncoplanar | ✔ | ✔ | ✘ | ✘ | $Ce_3NIn$ |
| | ✔ | ✘ | ✘ | ✔ | $MgV_2O_4$ |
| | ✘ | ✔ | ✔ | ✘ | CrSe |
| | ✘ | ✘ | ✔ | ✔ | $CuB_2O_4$ |

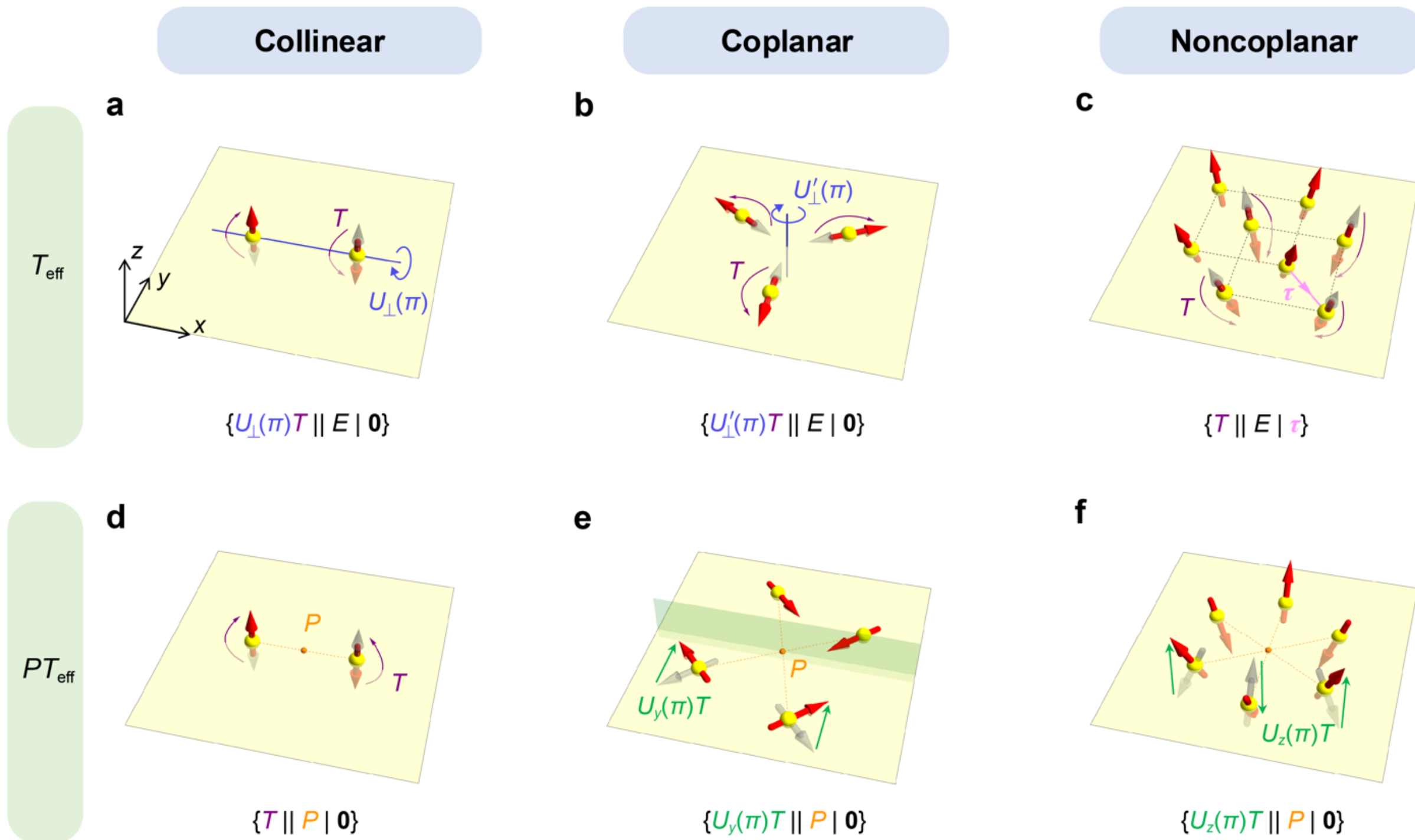

**Fig. 1 | Effective time-reversal symmetry. a-c**, Effective time-reversal symmetries always emerge in collinear (**a**) and coplanar AFMs (**b**), with $U_{\perp}(\pi)T$ and $U'_{\perp}(\pi)T$, respectively, and may emerge in certain noncoplanar AFMs (**c**) with $T\boldsymbol{\tau}$. **d-f**, Combined symmetry of spatial inversion and time-reversal $PT$ is available in certain collinear AFMs (**d**), while effective combined symmetry can emerge in coplanar (**e**) and noncoplanar AFMs (**f**) with $U_{\mathbf{n}}(\pi)TP$. $T$: time-reversal; $P$: spatial inversion; $U_{\perp}(\pi)$: two-fold spin rotation along an axis normal to the Néel vector; $U'_{\perp}(\pi)$: two-fold spin rotation along the axis normal to all in-plane magnetic moments; $U_{\mathbf{n}}(\pi)$: two-fold spin rotation along **n** axis; $\boldsymbol{\tau}$: fractional lattice translation. Red arrows and yellow balls denote magnetic moments and atoms, respectively, and shadowed arrows represent the intermediate state of magnetic moments operated by part of the combined symmetry, *i.e.* $T$ in **a-c** and $P$ in **d-f**.

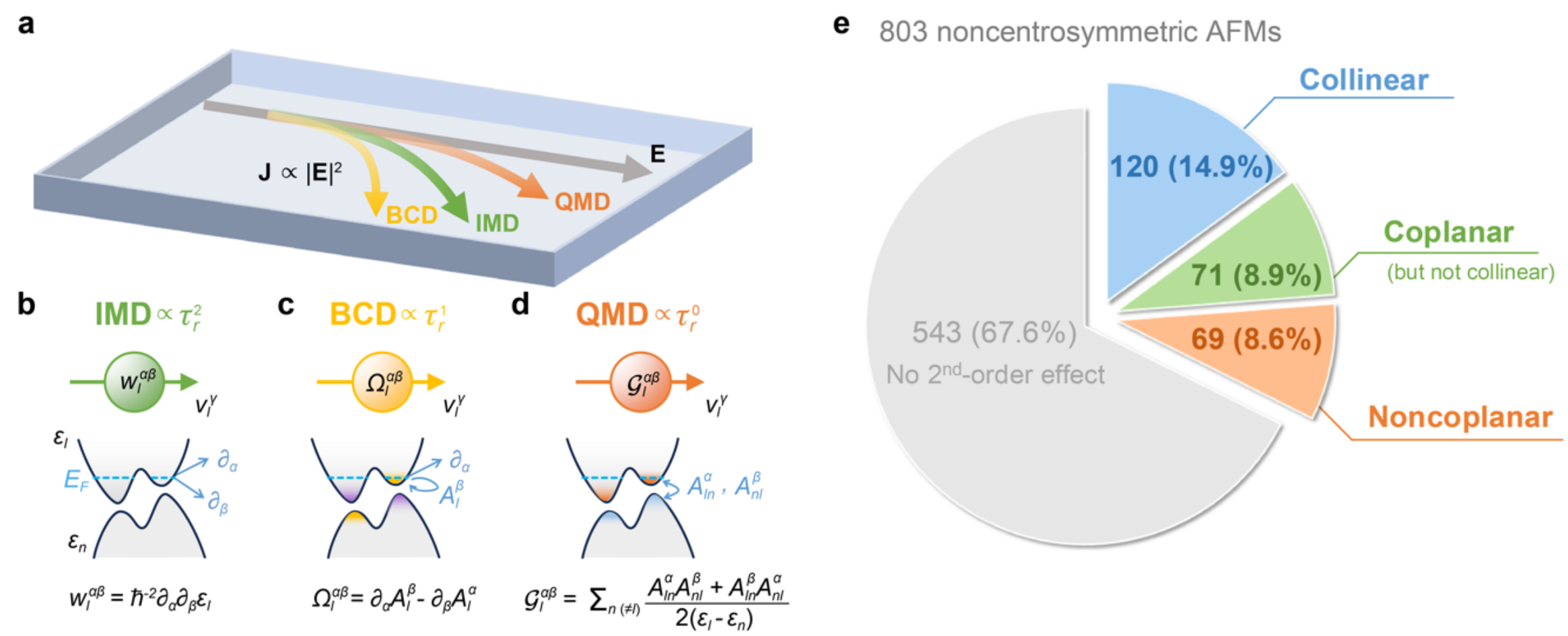

**Fig. 2 | Second-order transports and statistics of AFM material database. a**, Schematic of nonlinear charge current $\mathbf{J} \propto |\mathbf{E}|^2$ driven by the electric field **E**, where the current is contributed by three geometric quantities as IMD, BCD, and QMD. **b-d**, Physical mechanisms and formulae for IMD (**b**), BCD (**d**), and QMD (**d**), and their polynomial dependence on relaxation time $\tau_r$. **e**, Among 803 noncentrosymmetric AFMs, 543 AFMs have no second-order transport effects. Among the remaining 260 materials, 120 and 71 AFMs with collinear and coplanar antiferromagnetic geometry, respectively, are found to have BCD-contributed second-order transport effect, while 69 AFMs with noncoplanar antiferromagnetic geometry are found to allow second-order transport effect contributed by at least one geometry quantity. IMD: inverse mass dipole, BCD: Berry curvature dipole, QMD: quantum metric dipole.

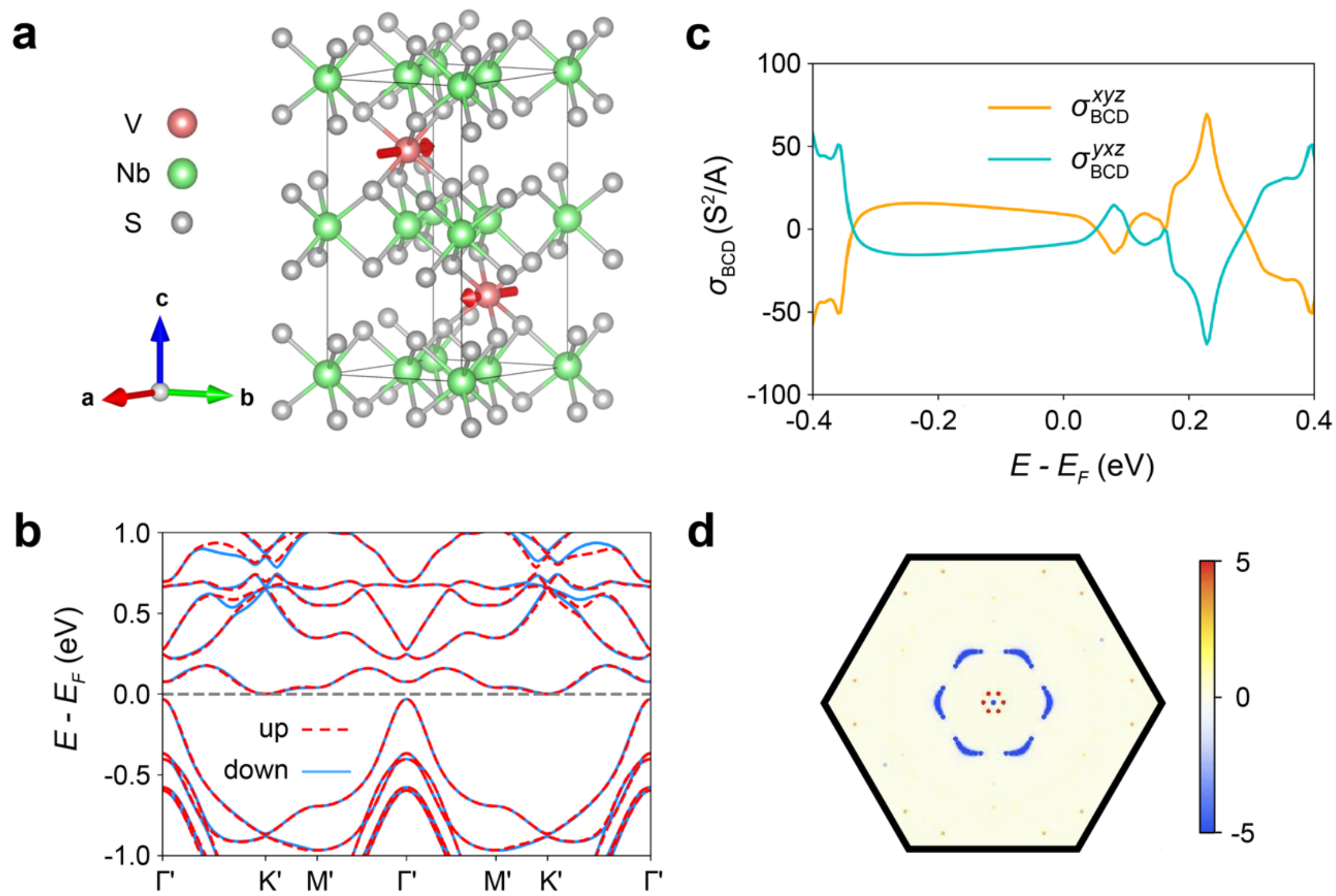

**Fig. 3 | Collinear AFM $VNb_3S_6$. a**, Crystal structure and collinear magnetic geometry of $VNb_3S_6$. **b**, DFT-calculated band structures with the projection onto opposite spin components. Spin-orbit coupling is turned off. **c**, Nonlinear conductivity tensor contributed by BCD with relaxation time set $\tau_r = 1$ ns. **d**, Distribution of the BCD $\partial^z \Omega^{xy}(k_x, k_y)$ in the slice of Brillouin zone $k_z = 0$ at $\sim$0.25 eV above the Fermi energy.

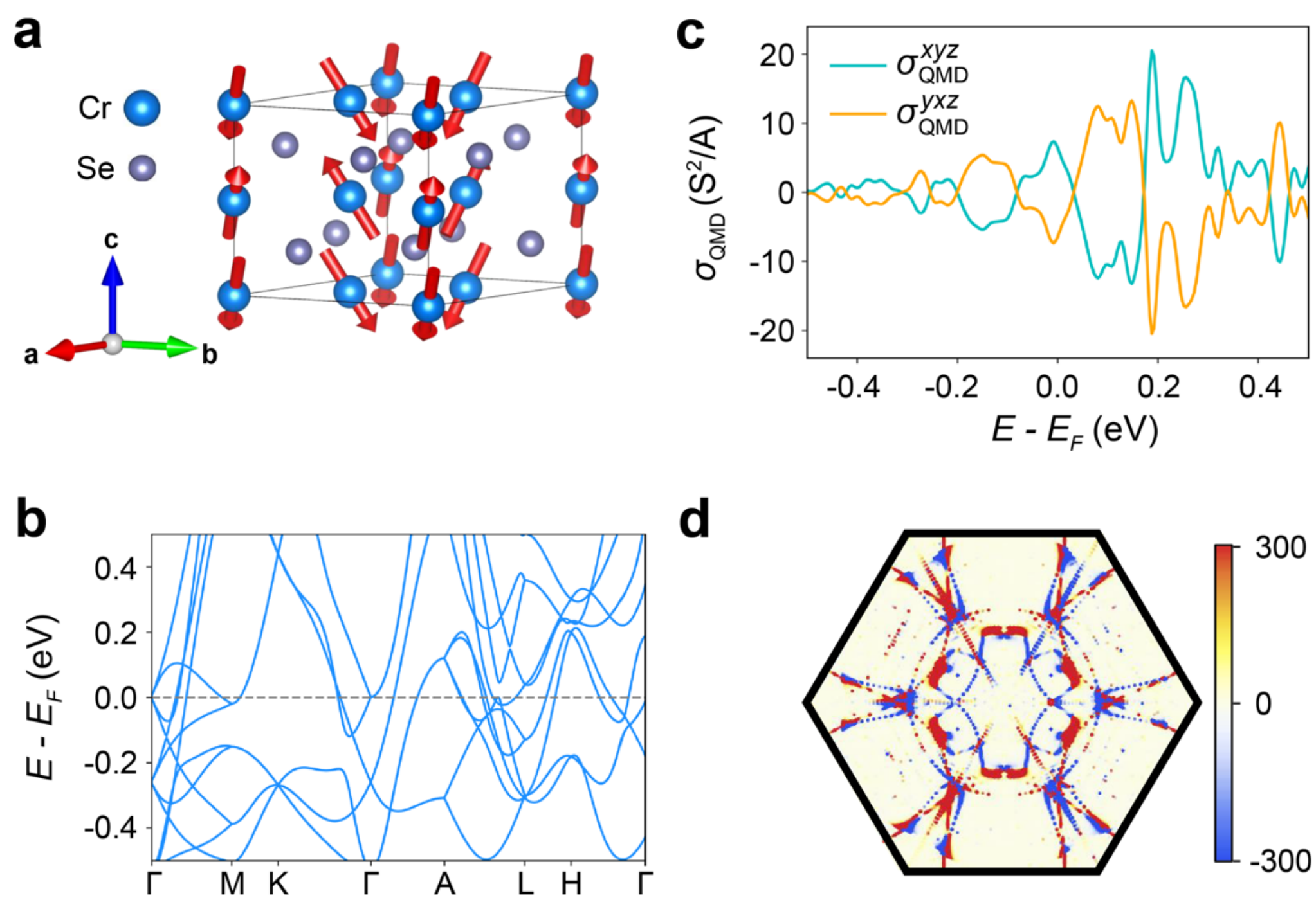

**Fig. 4 | Noncoplanar AFM CrSe. a**, Crystal structure and noncoplanar magnetic geometry of CrSe. **b**, DFT-calculated band structures without spin-orbit coupling. **c**, Nonlinear conductivity tensor contributed by QMD. **d**, Distribution of the QMD $\partial^x \mathcal{G}^{yz}(k_x, k_y)$ in $k_z = 0$ plane at $\sim$0.19 eV above Fermi energy.